\documentclass[prc,aps,nofootinbib,superscriptaddress,showkeys,showpacs,twocolumn,10pt,floatfix]{revtex4-1}
\usepackage{epsfig}
\usepackage{graphicx}
\usepackage{amsmath,amssymb,amsfonts}
\usepackage[small]{subfigure}

\usepackage[usenames]{color}
\usepackage{ulem}

\begin{document}
\title{Application of an \textit{ab-initio}-inspired energy density functional to nuclei: impact of the effective mass and the slope of the symmetry energy on bulk and surface properties.}

\author{Stefano Burrello}
\email{stefano.burrello@ijclab.in2p3.fr}
\affiliation{Universit{\'e} Paris-Saclay, CNRS/IN2P3, IJCLab, 91405 Orsay, France}
\author{J{\'e}r{\'e}my Bonnard}
\email{jeremy.bonnard@york.ac.uk}
\affiliation{Department of Physics, University of York, Heslington, York YO10 5DD, United Kingdom}
\author{Marcella Grasso}
\email{marcella.grasso@ijclab.in2p3.fr}
\affiliation{Universit{\'e} Paris-Saclay, CNRS/IN2P3, IJCLab, 91405 Orsay, France}

\begin{abstract}
The YGLO (Yang-Grasso-Lacroix-Orsay) functional is applied for the first time to investigate ground-state properties of different isotopic chains, from Oxygen to Lead. Mean-field Hartree-Fock calculations 
are carried out to analyze global trends for separation energies, binding energies, radii, neutron skins, and density profiles.
We have three objectives:
i) we study whether this functional leads to a reasonable description of ground-state properties (despite the fact that it was not adjusted on nuclei) and we discuss the associated limitations; ii) we investigate whether the correct description of the low-density nuclear gas, which is the peculiarity of this functional,  has any relevant impact on predictions for nuclei; iii) we connect nuclear energies, radii and density profiles with properties of the corresponding equations of state of infinite matter. In particular, we identify a link existing between the isoscalar effective mass 
and spatial properties in neutron-deficient nuclei, namely proton radii and tails of proton densities. On the other side, we show that the slope of the symmetry energy is connected with spatial properties in neutron-rich nuclei: the slope computed at saturation density is related to 
neutron skin thicknesses, as already well known, whereas the slope calculated at lower densities is linked to  the 
tails of neutron densities.   

The YGLO effective mass turns out to be quite low. Directions to improve this aspect are explored and suggested at the end of the manuscript.  
\end{abstract}
\date{\today}
\maketitle
%
\section{Introduction}
%
Possible connections between energy-density-functional (EDF) theories and effective-field theories (EFTs) are 
explored in the literature (see, for instance Refs. \cite{f1,fur,ppnp}). 
As a matter of fact, an underlying EFT is likely to exist in EDF theories, representing a microscopic basis for such phenomenological approaches. The reason for this conjecture is the global and well-known success met by 
EDF theories over the entire nuclear chart in both structure analyses and reaction applications.  
Some illustrations of these studies and proposed strategies, for example, to go towards the construction of a power counting in EDF are provided in Refs. \cite{yang2017,burrello2020}. These general reflections are carried out in a domain which is still poorly explored. 

On the other hand, the dilute regime of nuclear matter represents  
a specific sector of nuclear physics where an EFT and a corresponding small parameter governing an order-by-order expansion are well established, with an associated power counting. 
Starting from this, two types of EDFs, inspired by EFTs and benchmarked on {\it{ab-initio}} results, were recently proposed, YGLO (Yang-Grasso-Lacroix-Orsay) and ELYO (extended Lee-Yang
Orsay) \cite{yglo,elyo,drop1,drop2}. 
Both EDFs aim at accommodating two density regimes within a unique theoretical scheme. One, close to the equilibrium point of symmetric matter (SM), is the regime  where the parameters of phenomenological effective interactions are currently adjusted and is of interest for finite nuclei and for almost all their phenomenology.  The other one represents the limit of a dilute Fermi gas, described by the so-called Lee-Yang expansion (LYE). We remind that the first terms of such a low-density expansion were initially derived several decades ago by different authors (see for instance Refs. \cite{LY,HY,efimov,baker,bishop,ya}) and were revisited afterwards by applying EFT methods \cite{HF}. Very recently, also the fourth-order contribution was computed in EFT \cite{sc}.   
  
The YGLO functional has a hybrid form which contains a resummed formula together with Skyrme-type gradient and density-dependent terms. Resummation techniques are perfectly adapted to deal with both density regimes by employing a unique functional. As a matter of fact, they were introduced in EFT for handling systems characterized by extremely large values of the scattering length \cite{schafer,kaiser,steele}, as is the case, for example, in pure neutrom matter (NM) (the neutron-neutron $s$-wave scattering length being equal to $a_0=-18.9$ fm). 

In Ref. \cite{yglo}, a resummed formula allowed for properly including the first two terms of the LYE in the equation of state (EOS) of infinite matter. 
 Such a formula was adopted not only for the EOS of NM but also for the case of SM, with obvious different factors related to the different spin-isospin degeneracy in the two cases \cite{FW}. This was done with the aim of tackling SM and NM with the same type of functional, despite the fact that clusterization phenomena become relevant at low densities in SM \cite{bur15}, which are obviously not predicted within the proposed scheme. 

To describe clusterization phenomena in low-density SM, one should overcome the simple mean-field approximation that was adopted in Ref. \cite{yglo},  
by explicitly 
breaking the translational invariance. 
 Since a simple mean-field approach was employed in Ref. \cite{yglo}, the 
used low-density reference for SM was
 chosen as an EOS where such a symmetry is indeed not broken. Nevertheless, nothing prevents the YGLO functional from describing low-density clustering effects in matter 
 and in nuclei if the proper theoretical approach is used, analogously to 
what happens for example with EDF Skyrme functionals \cite{sk1, sk2, vau}, if the translational invariance is removed in matter \cite{bur15,negele}. 
Traditional EDF functionals have also predicted clustering effects at very low densities in nuclei, employing 3D calculations (see, for example, Ref. \cite{girod} based on Gogny functionals 
\cite{gogny1,gogny2} or 3D stochastic mean-field calculations in Ref. \cite{guarnera}). 

By construction, in Ref. \cite{yglo}, the very low-density regime was  described in NM as the sum of the first terms of an expansion in $(a_0 k_F)$, 
where $a_0$ is the neutron-neutron $s$-wave scattering length and $k_F$ is the Fermi momentum, related to the density $\rho$ by the relation $k_F= (3 \pi^2 \rho)^{1/3}$. The same was done for SM, where the $s$-wave scattering length $a_1$ was taken equal to the average in the spin-singlet $^1S_0$ channel, that is 
$a_1=-20.0$ fm (the spin-triplet $^3S_1$ neutron-proton contribution was neglected \cite{FW}). The relation between the Fermi momentum and the density is in this case given by $k_F= (3/2 \pi^2 \rho)^{1/3}$.  

Two parameters in the resummed formula were constrained by $a_0$ and $a_0^2$, to recover the linear and second-order terms of the LYE. For NM, the other parameters were benchmarked on QMC AV4 results \cite{geze} up to $|a_0 k_F| \approx$ 10 and, at higher densities, either on Akmal et al. results \cite{akmal} or on Friedman-Pandharipande (FP) results \cite{FP}, generating two versions of YGLO, YGLO(Akmal) and YGLO(FP), respectively. By construction, the NM EOS corresponding to the former is stiffer than the other. For SM, since the EOSs provided in Refs. \cite{akmal} and \cite{FP} are very similar, the FP EOS was chosen as unique benchmark.    
 
On the other side, the ELYO functional of Ref. \cite{elyo} was conceived to produce a NM EOS equal to  
the sum of the first terms of the LYE, without resummation.  
To guarantee that such a NM EOS is correct at all density scales (both very low densities and close to the equilibrium point of SM) a tuned, 
density-dependent neutron-neutron $s$-wave scattering length was 
proposed, satisfying the condition to converge to the known physical value of $-18.9$ fm in the very dilute regime.
In a first study, only $s$-wave terms of the LYE were considered \cite{elyo} and the expression for the NM EOS was truncated before the first term containing the $p$-wave scattering length. 
The functional was enriched more recently in Ref. \cite{drop2} by including also such a term.

Having introduced the functionals in infinite matter \cite{yglo,elyo}, a first attempt to go towards finite-size systems was discussed in Refs. \cite{drop1,drop2}, where applications to spherical neutron drops trapped in a harmonic potential were presented. For the YGLO case, this was done only for the YGLO(FP) version of the functional.  
We extend in this work also the YGLO(Akmal) functional to neutron drops and provide a set of parameters for this case.

With the two versions of the YGLO functional, we present here the first applications to finite nuclei based on a mean-field approach, by exploring different isotopic chains with different masses, from Oxygen to Lead.  
These applications have different objectives. The first and most obvious one is to check whether this type of functionals describes reasonably well properties of finite nuclei, compared to 
more traditional 
Skyrme EDFs 
, which is not obvious since the YGLO functionals were not adjusted to reproduce selected observables in nuclei.
The second objective is to check whether the peculiarity of these functionals, that is the correct description of the low-density regime, has any impact on  predictions obtained for finite nuclei. Finally, the third objective is to perform a more global analysis by connecting some of the obtained results for nuclei with specific properties of the corresponding EOSs, for instance the slope of the symmetry energy and  the effective mass. 
Being the YGLO effective mass quite low, some reflections are dedicated to possible improvements of this aspect.   
 
To perform these first analyses, we treat only even-even nuclei and neglect the effects related to superfluidity in open-shell systems. This is justified by the fact that we aim at analyzing  global trends which are not expected to be sensibly affected by the inclusion of a pairing interaction. We thus perform Hartree-Fock (HF) calculations in spherical symmetry.

The manuscript is organized as follows. 
We summarize the main properties of the YGLO functional in Sec. \ref{fun}. 
We extend the YGLO(Akmal) functional to neutron drops in Sec. \ref{drop} and provide the corresponding parameters. Section \ref{energy} contains the predictions for binding energies and two-neutron separation energies of several isotopic chains.  We present in Sec. \ref{radii} the results for radii and  neutron skins and discuss the known links existing between neutron skins and the slope of the symmetry energy at saturation density. The tails of the radial density profiles are analyzed in Sec. \ref{tail} for some neutron-deficient and neutron-rich nuclei. Links between such tails and the isoscalar effective mass (the slope of the symmetry energy at low density) are identified and discussed in neutron-deficient (neutron-rich) systems. 
For neutron-deficient nuclei, also the differences between neutron and proton radii are found to be connected to the isoscalar effective mass. 
Finally, Sec. \ref{emass} contains an analysis of the YGLO effective mass.  Conclusions and perspectives are summarized in Sec. \ref{concl}.  

\section{YGLO functional} \label{fun}

The EOS introduced for nuclear matter in Ref. \cite{yglo} is written as
\begin{equation} \label{YGLOEoS}
   \dfrac{E}{A} = K_\beta(\rho) + Y_\beta(\rho) \rho + D_\beta \rho^{5/3} + F_\beta \rho^{1+\alpha},
\end{equation}
where $\alpha=0.7$, $\beta=0$ (1) corresponds to NM (SM), 
$K_\beta$ represents the kinetic contribution 
and $Y_\beta$ stands for the aforementioned resummed formula given by
\begin{equation}
   Y_\beta(\rho) = \dfrac{B_\beta}{1-R_\beta\rho^{1/3}+C_\beta\rho^{2/3}}.
\end{equation}
The parameters $B_\beta$ and $R_\beta$ are constrained by the first two terms of the LYE 
in both NM and SM cases,
\begin{equation} \label{YGLOld}
   \begin{split}
   & B_\beta = \dfrac{2\pi\hbar^2}{m}\dfrac{\nu-1}{\nu} a_\beta, \\
   & R_\beta = \dfrac{6}{35\pi}\left(\dfrac{6\pi^2}{\nu}\right)^{1/3}(11-2\ln2) a_\beta, 
   \end{split}
\end{equation}
$\nu=2$ (4) being the spin-isospin degeneracy for $\beta=0$ (1). The other 
six parameters $C_\beta$, $D_\beta$, and $F_\beta$ are  adjusted on benchmark microscopic values and $a_\beta$ 
are the corresponding $s$-wave scattering 
lenghts.

When a finite-size system has to be treated, the underlying functional needs to be written explicitly. This was done in the first applications to neutron drops carried out with a Hartree-Fock-Bogoliubov (HFB) model in spherical symmetry \cite{drop1}.
In this case, one may write the energy density $\varepsilon$ as 
\begin{equation} 
   \varepsilon(\vec{r}) = \mathcal{T}(\vec{r}) +\varepsilon_\mathrm{\omega}(\vec{r})+ \varepsilon_\mathrm{c}(\vec{r}) + \varepsilon_\mathrm{so}(\vec{r}) + \varepsilon_\mathrm{pp}(\vec{r}),  \label{edffull}
\end{equation}
where one recognizes the kinetic term   
$\mathcal{T}(\vec{r})=\hbar^2\tau(\vec{r})/2m$, the trap contribution $\varepsilon_\mathrm{\omega}(\vec{r})$ corresponding to the harmonic potential $m\omega^2\vec{r}^{\,2}/2$,  
 the spin--orbit contribution $\varepsilon_\mathrm{so}(\vec{r})$, and the pairing one $\varepsilon_\mathrm{pp}(\vec{r})$. The latter was produced by a mixed surface-volume pairing interaction, 
\begin{equation}
   V(\vec{r}) = V_\mathrm{pp} \left( 1-\dfrac{1}{2}\dfrac{\rho(\vec{r})}{\rho_c}\right)\delta(\vec{r}),
\end{equation}
with $\rho_c=0.16$ fm${}^{-3}$. A cutoff of 60 MeV was taken in quasiparticle energies with a smooth diffuseness of 1 MeV.
The 
central contribution $\varepsilon_\mathrm{c}$ contains the terms generated by the YGLO functional itself. It is equal to 
\begin{equation} \label{YGLO2}
\begin{split}
   \varepsilon_c=&\bigl(2Y_1[\rho]-Y_0[\rho]\bigl)\rho^2 -2\bigl(Y_1[\rho]-Y_0[\rho]\bigr)\bigl(\rho_n^2+\rho_p^2\bigr) \\
       &+ \varepsilon_1 + \varepsilon_3 + \varepsilon_{3'},
       \end{split}
\end{equation}
where the isospin dependence 
on the first line, suggested in Ref. \cite{drop1}, ensues from the parabolic approximation employed in Ref. \cite{yglo} for the EOS of asymmetric matter. 
$\varepsilon_3$ corresponds to the term of the EOS containing the parameter $F_{\beta}$. Setting up a correspondence with a Skyrme-type functional, such a 
term is related to the density-dependent contribution [($t_3,x_3$) term], that implies the following relations:
\begin{equation} \label{t3YG}
\begin{split}
   &t_3=16F_1, \\
   &t_3(1-x_3)=24F_0.
\end{split}
\end{equation}
$\varepsilon_1$ and $\varepsilon_{3'}$ derive from the EOS contribution containing the parameter $D_{\beta}$, which is splitted into a second 
density-dependent term and a gradient part (in a term-by-term correspondence with a Skyrme functional), with the introduction of a splitting parameter $W$. For the density-dependent contribution it holds  
\begin{equation} \label{t32YG}
\begin{split}
   &t_{3'}=16(1-W)D_1, \\
   &t_{3'}(1-x_{3'})=24(1-W)D_0.
\end{split}
\end{equation}
Regarding the gradient term, the choice of taking $t_2=x_2=0$ was adopted in Ref. \cite{drop1}, which implies 
 \begin{equation} \label{t1YG2}
\begin{split}
  & t_1=\dfrac{80}{9}W \left(\dfrac{3\pi^2}{2}\right)^{-2/3}D_1, \\
  & t_1(1-x_1)=\dfrac{40}{3}W (3\pi^2)^{-2/3}D_0.
\end{split}
\end{equation}
With this, the functional is ready to be applied to finite systems. The numerical values of the parameters $C_{\beta}$, $D_{\beta}$ and $F_{\beta}$ were reported in Ref. \cite{yglo} for both YGLO(FP) and YGLO(Akmal). The parameters related to the extension to neutron drops (splitting parameter $W$, spin-orbit coupling constant $V_{so}$ and pairing intensity $V_{pp}$) were listed in Ref. \cite{drop1} for the case of YGLO(FP).  
Throughout the following discussions, it is worth keeping in mind that the YGLO EDFs are less phenomenological than conventional Skyrme functionals in the sense that they involve one adjustable parameter less.

\section{YGLO(Akmal) and YGLO(FP) in neutron drops} \label{drop}
%
\begin{figure}
\begin{center}
\includegraphics[width=\columnwidth]{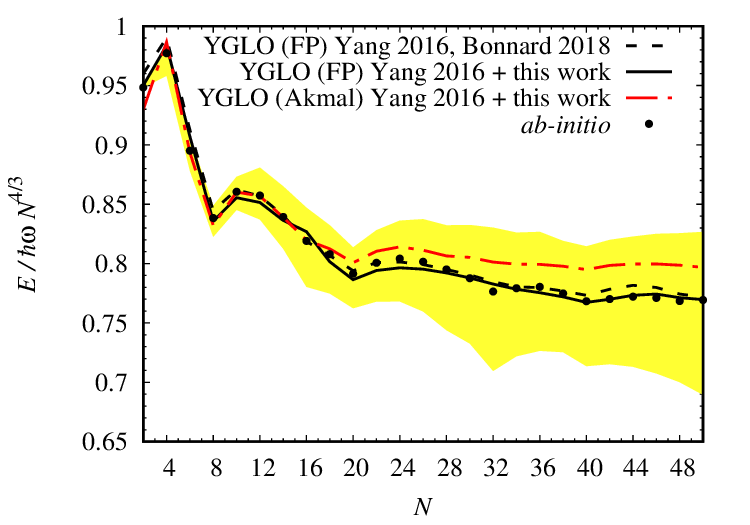}
\vspace*{-0.7cm}
\caption{
Droplet energies scaled by the value obtained in the Thomas-Fermi approximation as a function of the number of neutrons $N$, in a trap of frequency $\hbar \omega =$ 10 MeV. The yellow area represents the region where the reference {\it{ab-initio}} results are located (see text). The average of these reference values is indicated by the black circles. The red dot--dashed line corresponds to the values obtained with YGLO(Akmal). The black dashed line corresponds to the results of Ref. \cite{drop1}. The black solid line is obtained with the new FP set introduced here. Yang 2016: Ref. \cite{yglo}, Bonnard 2018: Ref. \cite{drop1}.}
\label{dr}
\end{center}
\end{figure}

For the adjustment of the parameters, we have followed the same fitting procedure as the one adopted for the YGLO(FP) case in Ref. 
\cite{drop1}. The splitting parameter $W$, the pairing strength $V_\mathrm{pp}$ and the spin-orbit coupling constant $V_\mathrm{so}$ are adjusted to reproduce a benchmark {\it{ab-initio}} curve, for the trap frequency $\hbar \omega=$ 10 MeV, by performing HFB calculations. The benchmark curve is  chosen, following Ref. \cite{drop1}, as the average of several microscopic results \cite{maris,gandolfi,potter}. The adjustment is done for neutron numbers $N$ ranging from 6 to 20, where the microscopic predictions are still located in a relatively narrow area around the average curve and the number of neutrons is large enough to guarantee that mean-field calculations may be regarded as trustworthy. For larger values of $N$, the dispersion of the microscopic results around the average becomes sensibly more important.
 
The energies of the drops, scaled by the values obtained in the Thomas-Fermi approximation, 
for even neutron numbers, are displayed on Fig.\ref{dr}. 
The yellow area contains the reference {\it{ab-initio}} results \cite{maris,gandolfi,potter} and the black circles are the average points adopted as benchmark values.   
The YGLO(Akmal) results are displayed with a red dot-dashed line and the corresponding values of the parameters are reported on Table I. 
For comparison, also the YGLO(FP) results published in Refs. \cite{drop1} are shown in the figure (black dashed line) and the corresponding parameters are recalled in Table I. 
One observes that beyond the region of the fit ($N>$ 20), 
both parametrizations of YGLO provide 
energies lying well inside the range defined by the sample of ab-initio results. Nonetheless,
the agreement 
with the benchmark values is less satisfactory for YGLO(Akmal) than for the case of YGLO(FP).  

A new parametrization for YGLO(FP) was also produced in this work, leading to energies which are slightly closer to the benchmark values, at least in the region $N>40$. The corresponding curve is also plotted in the figure (black solid line) and the parameters are reported on Table I. The spin-orbit strength differs very little from the original value provided in Ref. \cite{drop1} while the pairing coupling constant is reduced. The predictions obtained with the two FP sets do not present significant  differences here, but differences become more important when nuclei are considered, as we will see in Sec. \ref{energy} for the binding energies. 

Let us note that the relative 
position of the YGLO(FP) and YGLO(Akmal) 
curves on Fig. \ref{dr} is consistent with the microscopic NM calculations on which they have respectively been fitted. The Akmal et al. EOS \cite{akmal} is indeed stiffer that the FP one \cite{FP}. As we shall see below, this behavior remains visible in the properties of neutron-rich nuclei.

The isoscalar and neutron effective masses, computed at the equilibrium density of SM, are reported on Table II for the three sets. One observes that these values are quite low and this will be discussed later in the article. 

In spite of the fact that the functional was adjusted with HFB calculations to both open- and closed-shell drops, we neglect pairing correlations in the applications done here to nuclei. This corresponds to using a set of parameters with the same coupling constants but $V_\mathrm{pp}=0$. Such a set of parameters is able by construction to describe shell-closed systems in neutron drops and is thus well adapted for performing HF calculations. 

In what follows, we carry out HF calculations in spherical symmetry and in coordinate space for finite nuclei, with a box radius equal to 22 fm and a radial step of 0.1 fm. 

\begin{widetext}
\begin{center}
\begin{table}[t]
\label{parame}
\caption{Parameters for the sets YGLO(FP) and YGLO(Akmal). In all cases $\alpha=$ 0.7.}
\begin{tabular}{ccccccc}
\hline
 & $C_{\beta=0(1)} $ &	  $D_{\beta=0(1)} $   &  $F_{\beta=0(1)} $   &     W     & $V_\mathrm{so}$ & $V_\mathrm{pp}$ \\
 &          (fm$^2$)  &               (MeV fm$^5$)    &       (MeV fm$^{3+3\alpha}$) &   &    (MeV fm$^{5}$)   &    (MeV fm$^{3}$) \\
\hline      
YGLO(FP) \cite{yglo,drop1} & 100.87 (8.188) & -9264.18 (-6624.87) & 9571.90 (6995.46)   &  -0.0840  &   	      138.2          & 	       -275.1       \\ 
YGLO(FP) \cite{yglo} + this work & 100.87 (8.188) & -9264.18 (-6624.87) & 9571.90 (6995.46)   &  -0.0764 &   	      134.2          & 	       -154.7       \\  
YGLO(Akmal) \cite{yglo} + this work & 70.19 (8.188) & -8377.83 (-6624.87) & 8743.85 (6995.46)   &  -0.0680  &   	      130.0          & 	       -102.2         \\
\hline
\end{tabular}%
\end{table}
\end{center}
\end{widetext}

\begin{center}
\begin{table}[t]
\label{effe}
\caption{Neutron and isoscalar effective masses  computed at the saturation density of SM for the sets YGLO(FP) and YGLO(Akmal). The saturation density of SM is equal to 0.168 fm$^{-3}$ 
in all cases by construction.}
\begin{tabular}{ccc}
\hline
 & $(m^*/m)_s $ & $(m^*/m)_n$  \\
\hline      
YGLO(FP) \cite{yglo,drop1} & 0.445 & 0.476     \\ 
YGLO(FP) \cite{yglo} + this work & 0.468 & 0.500     \\  
YGLO(Akmal) \cite{yglo} + this work &   0.497 & 0.554        \\
\hline
\end{tabular}%
\end{table}
\end{center}

%
%
\section{Separation and binding energies. Links with the equation of state and the effective mass} \label{energy}
%

\begin{figure*}[t]
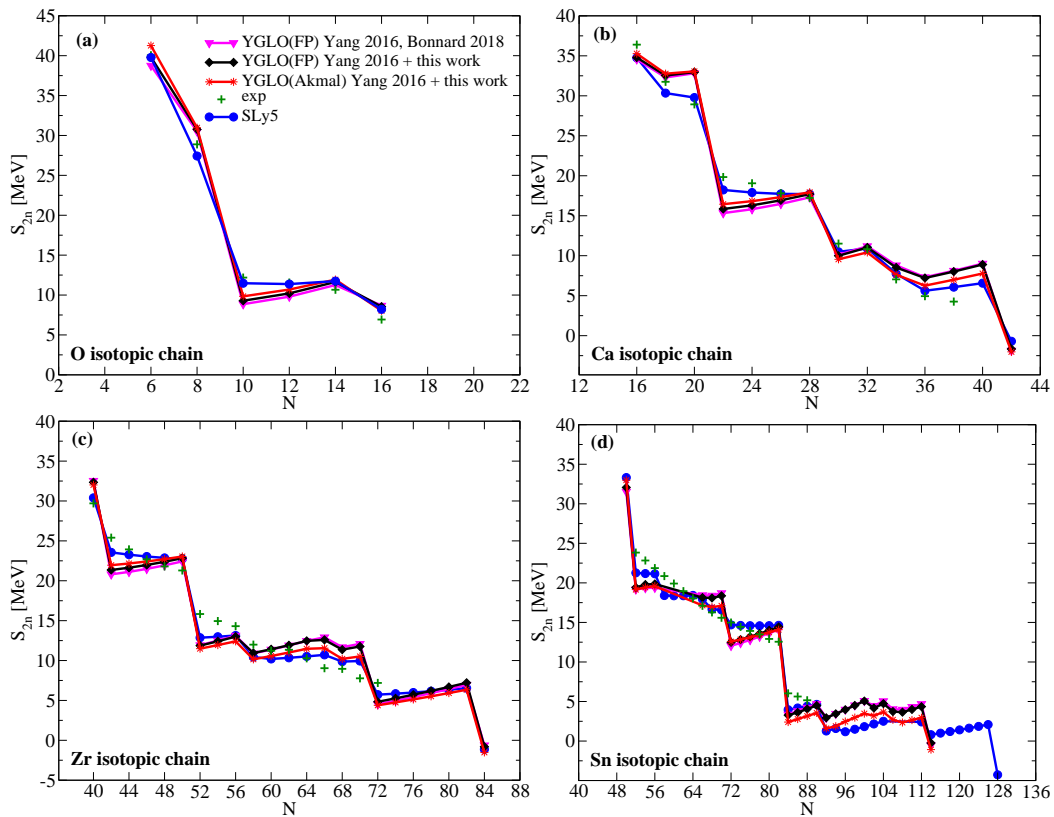

\begin{center}
\includegraphics[width=0.8\columnwidth]{O_S2N.eps}\includegraphics[width=0.8\columnwidth]{Ca_S2N.eps}\\ 
\includegraphics[width=0.8\columnwidth]{Zr_S2N.eps}\includegraphics[width=0.8\columnwidth]{Sn_S2N.eps}\\
\caption{Two-neutron separation energies $S_{2n}$ as a function of the neutron number $N$, along the O (a), Ca (b), Zr (c), and Sn (d) 
isotopic chains. The results corresponding to the different parameterizations of the YGLO functional considered in our study are compared with the ones obtained with the SLy5 interaction. The experimental data extracted from the National Nuclear Data Center \citep{nndc} are also reported, whenever available. Yang 2016: Ref. \cite{yglo}, Bonnard 2018: Ref. \cite{drop1}.}
\label{s2n}
\end{center}
\end{figure*}

As an illustration, we show in Fig. \ref{s2n} the two-neutron separation energies $S_{2n}=B(Z,N)-B(Z,N-2)$ for four 
selected isotopic chains, namely O, Ca, Zr, and Sn, where $B$ is the binding energy. The results are compared with the HF predictions obtained with the SLy5 \cite{sly5} Skyrme interaction and 
with the corresponding experimental data, wherever available. 
For the case of Oxygen in panel (a), we show the results up to the experimentally known drip-line nucleus, $^{24}$O. 
For the other isotopic chains, we plot the theoretical values up to the predicted drip-line position in each case, that is up to the last isotope where the separation energy is still positive. 

\begin{figure*}[t]
\begin{center}
\includegraphics[width=0.9\columnwidth]{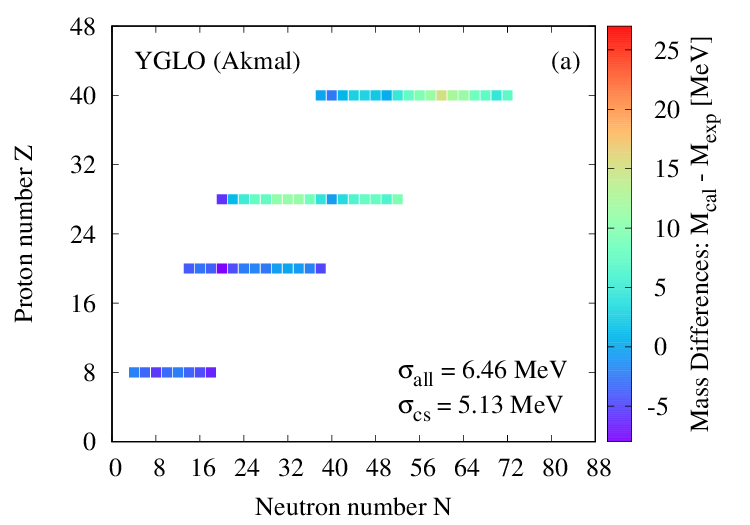}\includegraphics[width=0.9\columnwidth]{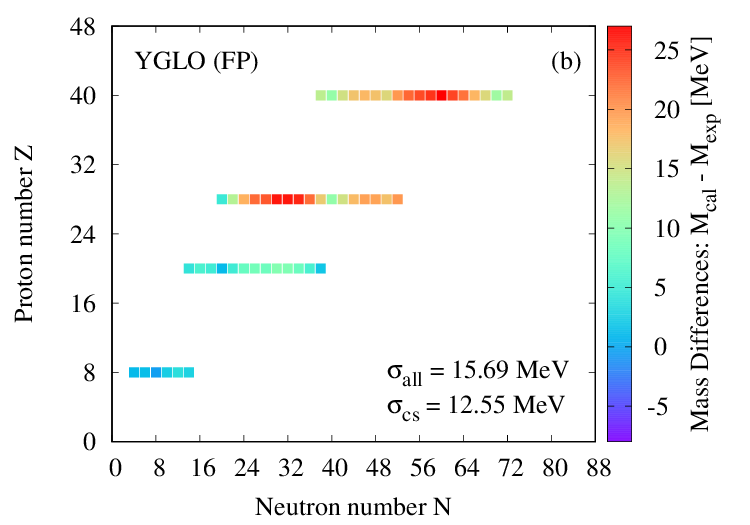}
\caption{(a) Differences between the HF binding energies obtained with the YGLO(Akmal) parameterization and the correponding experimental values extracted from Ref. \citep{nndc}. 
Two values of root mean square deviations are indicated in the figure: $\sigma_{all}$ and $\sigma_{cs}$, evaluated by taking into account all the considered nuclei of the isotopic chains or limiting only to closed-shell isotopes, respectively; (b) Same as in (a) but with the YGLO(FP) set of parameters (\cite{yglo} + this work) reported in Table I.}
\label{massd}
\end{center}
\end{figure*}

The figure clearly shows that the three versions of the YGLO functional lead to very similar $S_{2n}$ values, although some differences occur when moving towards the neutron drip-line, especially in the case of heavier systems. These values are close to the SLy5 results. One may also observe a reasonable global agreement with the shown experimental data. This agreement is found to be better for lighter systems, namely in Oxygen and Calcium isotopic chains. 

We remind that, since pairing correlations are neglected here, all the HF results (both YGLO and SLy5) are not expected to properly reproduce the experimental trends between two successive shell closures, where the experimental separation energies decrease monotonically. In addition, owing to the very low value of the YGLO effective mass (that will be discussed later in the article) the level spacing around the Fermi energy produced by the YGLO functionals is expected to be very large (very low level density around the Fermi energy). For this reason, 
the YGLO separation energies show shell closures which are much more pronounced compared to the SLy5 case. This is clearly visible in the figure and is indicated by the stronger reduction of the predicted YGLO $S_{2n}$ value just after a shell-closed nucleus. Such a  decrease is more important than in the Skyrme case and  is followed by 
a systematic increase of the $S_{2n}$ values in the next isotopes, where the neutron level is filled up to the successive shell closure. Such a behavior, which is in direct opposition with the expectation coming from the experiments, is less visible in the SLy5 case, where one observes in most cases a flat evolution between two successive shell closures. 

\begin{figure*}[t]
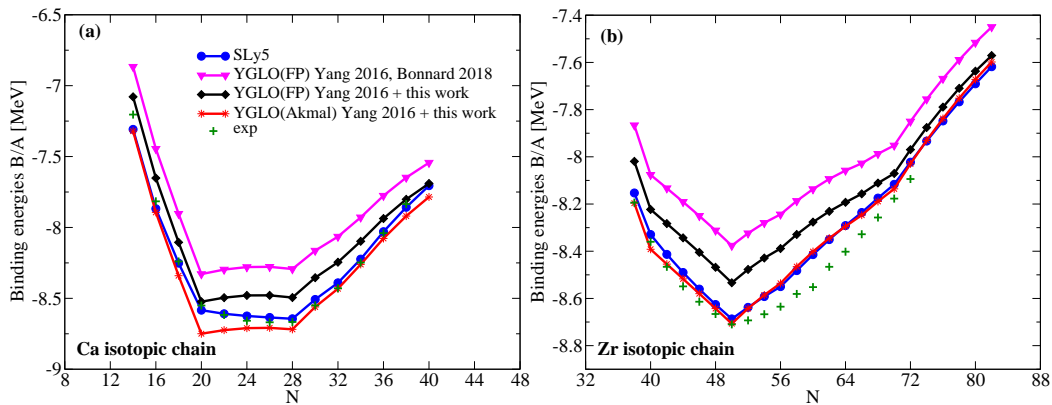

\begin{center}
\includegraphics[width=0.8\columnwidth]{Ca_bind.eps}\includegraphics[width=0.8\columnwidth]{Zr_bind.eps}
\caption{(a) Binding energies $B/A$ as a function of the neutron number $N$, along the Calcium isotopic chain. The results corresponding to the different parameterizations of the YGLO functional are compared with the ones obtained for the SLy5 Skyrme-like interaction. Experimental data from the National Nuclear Data Center \citep{nndc} are also reported, whenever available; (b) Same as in (a) but for Zr isotopes. Yang 2016: Ref. \cite{yglo}, Bonnard 2018: Ref. \cite{drop1}.} 
\label{be}
\end{center}
\end{figure*}
   
The analysis of the position of the two-neutron drip line in different isotopic chains can tell us if the YGLO correct description of very low densities is expected to induce any visible effect there, impacting in a  systematic way binding energies of exotic neutron-rich systems.  
We find that all the theoretical approaches lead to the same position of the two-neutron drip line in Ca and Zr isotopic chains, namely $^{60}$Ca and $^{122}$Zr, respectively.
In the case of Sn, the last nucleus for which the separation energy is positive is  $^{162}$Sn with the YGLO functionals and $^{176}$Sn with SLy5. 
For the other isotopic chains 
examined in this work, namely O, Ni and Pb, the drip line has been found either at the same position or shifted by a few isotopes compared to the Skyrme case. 
We remind however that also different Skyrme parametrizations may show similar discrepancies in the prediction of the drip line locations, which can be of 10 isotopes or even more for heavier systems. Thus, we may conclude that our results do not indicate any specific and 
systematic modifications of the neutron drip line predictions, that would be induced by the YGLO functionals, compared to more traditional interactions. 

The overall reasonable agreement that was found with the phenomenological functional SLy5 and with the available experimental values is certainly reassuring. 
However, separation energies are differences of binding energies. Such an agreement simply indicates that the variations of the binding energies along an isotopic chain are reasonable compared to the experimental data. But nothing guarantees that the absolute values of the binding energies are in good agreement with the experiment. We indeed expect that this is not the case. Experience teaches us that, in general, a tuning of parameters on some nuclei is necessary to get this goal (see for example the tuning on nuclei done for the KIDS functional in Ref. \cite{gil}). 

We have evaluated the differences between the mass values produced in our calculations and the corresponding experimental data and  have found in general quite large deviations. 
These differences are displayed in Fig. \ref{massd} for the sets YGLO(Akmal) and YGLO(FP) introduced in this work.
This is done for O, Ca, Ni, and Zr isotopes. Heavier isotopic chains are excluded  because deviations are found to be even larger in those cases.  In the plots, the root mean square deviations are also indicated, namely $\sigma_{all}$ and $\sigma_{cs}$, which are evaluated, respectively, by taking into account all nuclei of the isotopic chains shown in the figures or limiting only to closed-shell isotopes. 
A better agreement is of course expected for closed-shell nuclei because we neglect pairing correlations in our  calculations. The values of $\sigma$ confirm this expectation,  being $\sigma_{cs}$ smaller than $\sigma_{all}$ for the two cases. In general, deviations are smaller for the lightest isotopes, O and Ca for the Akmal case and O and some Ca isotopes for the FP case. Globally, the deviations corresponding to the Akmal functional (panel (a)) are sensibly smaller than those corresponding to the FP functional (panel (b)). In both cases, however,    
 a systematic deterioration arises when moving towards heavier and more isospin-asymmetric systems. 
It is worthwhile to notice that the level of accuracy reached in this restricted region of the nuclear chart by other mean-field HF models is not dramatically improved compared to the YGLO(Akmal) case. For example, in the case of HF calculations done with the SLy5 Skyrme parametrization, one obtains $\sigma_{all} =$ 4.75 MeV. If, however, only closed-shell nuclei are considered, $\sigma_{cs}$ is much smaller (1.96 MeV) in the SLy5 case. 

The observed discrepancies are after all comprehensible. Apart from the fact that, as already underlined,  we do not perform any fine tuning on selected nuclear binding energies, 
we should also recall that the adjustment of the parameters introduced for neutron drops was performed on the energy of drops with $N \le 20$. There, the available {\it{ab-initio}} data are quite similar and are located in a narrow region. 
We thus expect in general better predictions of experimental binding energies with our functionals for not too heavy nuclear systems. 
For heavier neutron drops, the available {\it{ab-initio}} results are more spread and it is not guaranteed  that the choice that we make for the benchmark (the average curve) is the one which is the most suitable for applications to nuclei. 
Indeed, deviations on the nuclear binding energies are found to be smaller in the Akmal case which, on the other hand, reproduces worse the benchmark energy curve in the drops with $N>20$ (Fig. 1). 
  
We remind also that the lack of the $p$-wave contribution in these versions of the YGLO functional may play a relevant role especially in the case of heavier systems, as already observed for neutron drops with the ELYO functional \cite{drop1,drop2}. 

Of course, absolute deviations have a different meaning in light than in heavy nuclei, being the total binding energy much larger (in absolute value) in heavy than in light systems.  
We cannot be too demanding by requiring absolute deviations comparable to those produced by models specifically adjusted on masses. A complementary information may come from relative deviations. We have found for instance that total binding energies differ from the experimental values by 4.3 (0.4) and 2.7 (0.7) \% in $^{16}$O and $^{24}$O, respectively, with the Akmal (FP) functional. The percentages are 2.3 (0.3), 0.6 (2.0), 1.0 (2.7), 0.03 (2.0) \% for $^{40}$Ca, $^{48}$Ca, $^{78}$Ni, and $^{90}$Zr, respectively, with Akmal (FP). This means that, even if precise absolute values cannot be predicted in such a model where mass constraints are missing, predictions are nevertheless not too far from the experimental data, with relative deviations smaller than 4\% in all cases considered above.   

Regardless the differences addressed above, our analysis with YGLO is also useful to highlight several other qualitative features. For example, we show in Fig. \ref{be} the binding energies $B/A$ for Ca and Zr isotopic chains. It is quite interesting to note that, despite the fact that the absolute values are shifted from one another, the Akmal and FP show a similar trend (the curves are almost parallel) when varying the neutron number, at least when limiting to almost isospin-symmetric nuclei. 
When moving towards more neutron-rich systems the two curves exhibit different slopes and the one corresponding to Akmal becomes more similar to the SLy5 one.
Such a result reflects the way on which YGLO functionals were built, since the FP and Akmal versions have exactly the same EOS for SM and differ only for the case of pure NM, where the Akmal EOS returns values which are rather close to the SLy5 ones. These features are in turn confirmed by the binding energy behavior versus $N$. 

We find that the YGLO(FP) \cite{yglo,drop1} functional provides less bound systems than the YGLO(FP) \cite{yglo} (+ this work) one, which in turns leads to less bound systems than the YGLO(Akmal) \cite{yglo} (+ this work) functional. 

Figure \ref{be} also shows that, globally, the absolute binding energies obtained with the  YGLO(FP) \cite{yglo} (+ this work) functional  are closer to the experimental values than those provided by the 
 YGLO(FP) \cite{yglo,drop1} functional. For this reason, we have employed only the YGLO(FP) \cite{yglo} (+ this work) version in Fig. \ref{massd} and we will systematically employ it in the following part of the manuscript. 

\begin{figure}[t]
\begin{center}
\includegraphics[width=\columnwidth]{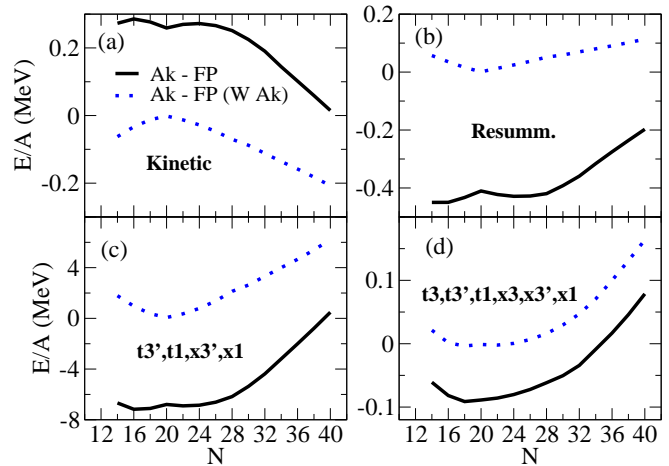}
\caption{Black solid lines: differences between YGLO(Akmal) and YGLO(FP) contributions to the total energy per nucleon along the Ca isotopic chain corresponding to: (a) the kinetic contribution, (b) the resummed contribution, (c) the sum of the contributions depending on the parameters $t_1$, $t_{3'}$, $x_1$, and $x_{3'}$, (d) the sum of the contributions depending on the parameters $t_1$, $t_3$, $t_{3'}$, $x_1$, $x_3$, and $x_{3'}$. The blue dotted lines correspond to the same differences obtained by using the $W$ parameter of the YGLO(Akmal) functional in the FP case.}
\label{diffe}
\end{center}
\end{figure}

To understand the reason for these global and systematic discrepancies in the total binding energies (in spite of the fact that all the YGLO functionals were adjusted on the same benchmark curve in neutron drops and on the same EOS of SM) we analyze separately, for Ca isotopes, some of the contributions to the total energy per nucleon. We show in Fig. \ref{diffe} the differences between the results obtained with the two functionals YGLO(Akmal) and YGLO(FP) for the following quantities: the kinetic contribution (a), the resummed contribution (b), the sum of the contributions depending on $t_1$, $x_1$, $t_3'$, and $x_3'$ (c), and the sum of the contributions depending on $t_1$, $x_1$, $t_3'$, $x_3'$, $t_3$, and $x_3$  (d). Two lines are shown in each panel: the black solid lines refer to the differences between the Akmal and the FP results, whereas the blue dotted lines represent the same differences, but having used the $W$ value of the Akmal functional in the FP case. 

Let us examine the case where the parameter $W$ is taken equal. The fact that $W$ is the same 
implies automatically that the two functionals have exactly the same 
isoscalar effective mass, but they still have different neutron effective masses because their $x_1$ parameters 
have different values 
(different $D_0$ in Eq. \eqref{t1YG2}). Nevertheless, in practice, effective mass values are very close. In light of the fact that the SM EOSs are the same, the $t_3$ parameters are equal, independently of $W$. If $W$ is the same, also the parameters $t_1$ and 
$t_3'$ are equal. 
As a result, the two YGLO functionals provide almost the same energy in the $N=Z$ nucleus $^{40}$Ca, as one can see in Fig. \ref{diffe}, where the blue dotted lines are practically zero at $N=20$. We verify in such a way that the discrepancies observed in Fig. \ref{be} between the total binding energies obtained with the two YGLO functionals merely depend on the choice of the splitting parameter, at least in the case of almost symmetric nuclei.
On the other hand, the trend exhibited by the blue dotted lines versus $N$ allows one to isolate the genuine effect of varying the NM EOS, while maintaining unaltered the other ingredients. For the NM EOSs it turns out that, at neutron densities beyond 0.05 fm$^{-3}$, the Akmal EOS becomes stiffer than the FP one, leading in such a way to less bound neutron matter.  When the neutron excess grows moving towards more neutron-rich systems 
(beyond $N=28$ in the figure) the discrepancies between Akmal and FP increase in all panels. The YGLO(Akmal) functional leads to less bound nuclei compared to FP and, when the isospin asymmetry increases, the difference between the two functionals progressively grows, which simply reflects the higher stiffness of the Akmal NM EOS.
This behavior guides us also in interpreting the differences between the energy contributions obtained using the optimal values of the splitting parameters $W$ for the two YGLO functionals (black solid lines of Fig. \ref{diffe}). 
The optimal choice of 
$W$ for FP seems to favor configurations which globally predict less bound systems with respect to the Akmal case, at least in the case of almost symmetric nuclei. 
The black curves are shifted (due to the different $W$ value in the FP case) compared to the blue dotted ones, but maintain approximately the same trends and slopes as in the other case. Owing to the higher stiffness of the Akmal NM EOS, 
 going towards more neutron-rich systems, the tendency is to attenuate the Akmal extra binding. In general, for neutron-rich systems, the higher stiffness of the Akmal NM EOS compared to the FP one tends to reduce the discrepancy between the two sets of binding energies, which become closer. Such a behavior is general and may be observed also in the trend of the total binding energies for the case of Zr isotopes (Fig. \ref{be}(b)).

\section{Radii and neutron skins. Links with the slope of the symmetry energy at saturation} \label{radii}
%

\begin{figure}
\begin{center}
\includegraphics[width=\columnwidth]{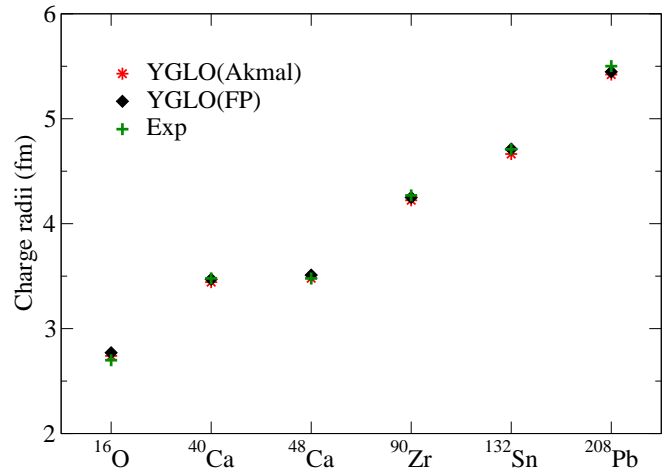}
\vspace*{-0.7cm}
\caption{Charge radii of some closed-shell nuclei computed with YGLO(Akmal) and YGLO(FP). The experimental values are extracted from Ref. 
\cite{raexp}.  }
\label{rad}
\end{center}
\end{figure}

Let us first compare the charge radii of some chosen closed-shell isotopes with the corresponding experimental values. As an illustration, we show in Fig. \ref{rad} the charge radii of $^{16}$O, $^{40}$Ca, $^{48}$Ca, $^{90}$Zr, $^{132}$Sn, and $^{208}$Pb. 

The YGLO values are almost superposed in the figure and in good agreement with the experimental data. We may thus conclude that, whereas a fine tuning on masses is probably required to correctly reproduce binding energies, as discussed in the previous section, nuclear radii are already globally well described with the available sets of parameters.

\begin{figure*}[t]
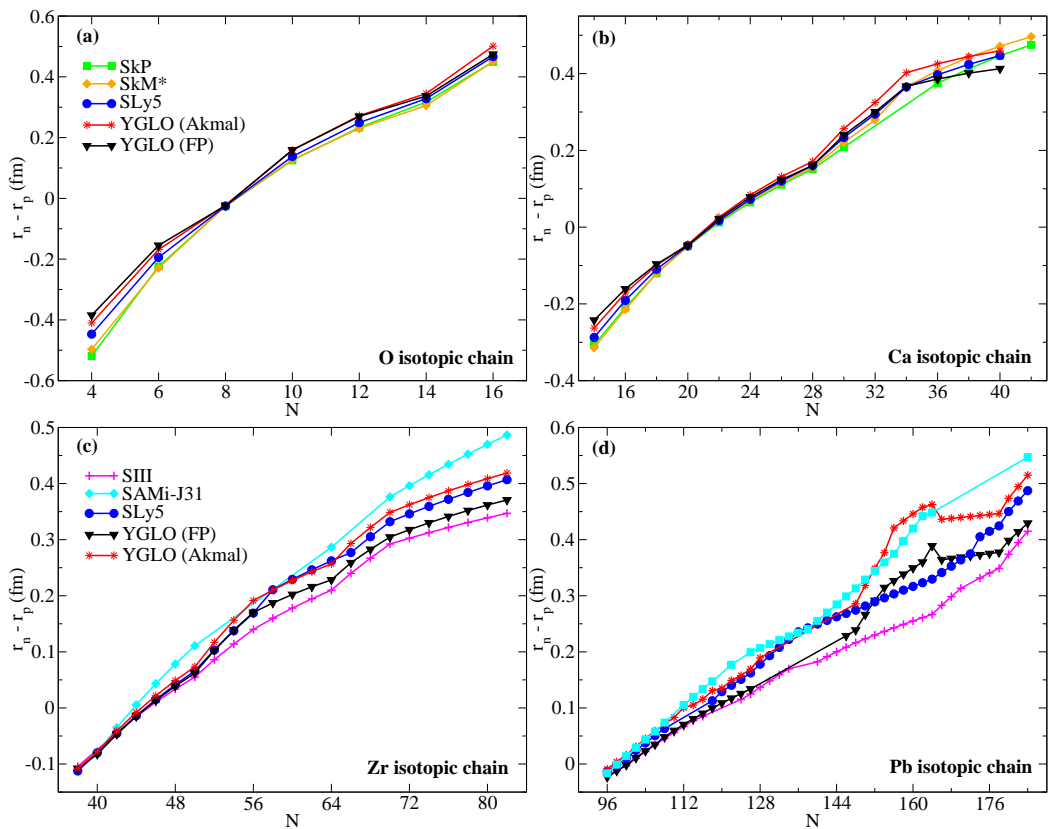

\begin{center}
\includegraphics[width=0.8\columnwidth]{O_skin_n_sly5_yglo_skm_siii_skp.eps}\includegraphics[width=0.8\columnwidth]{Ca_skin_n_sly5_yglo_skm_siii_skp.eps}\\ 
\includegraphics[width=0.8\columnwidth]{Zr_skin_n_sly5_yglo_siii_sami31.eps}\includegraphics[width=0.8\columnwidth]{Pb_skin_n_sly5_yglo_siii_sami31.eps}\\
\caption{Difference between neutron and proton radii for O (a), Ca (b), Zr (c), and Pb (d) isotopes. The legend of panel (a) applies also 
to panel (b). The legend of panel (c) applies also to panel (d).}
\label{skin}
\end{center}
\end{figure*}

It is interesting to look also at the evolution of the neutron and proton radii, when moving along a given isotopic chain. We plot in Fig. \ref{skin} the difference between the neutron and the proton radii $r_n-r_p$ for O (a), Ca (b), Zr (c), and Pb (d) isotopes, up to $^{24}$O for the O case and up to the predicted drip line nuclei in the other cases. The predictions obtained with the Akmal and FP versions of the YGLO functional are compared with the results deduced when adopting several standard EDFs, which are available from the literature. For these comparisons, we have chosen in particular the Skyrme parametrizations SLy5, SkM* \cite{skm}, SkP \cite{skp}, SIII \cite{siii}, and SAMi-J31 \cite{sami}.  In such a way, employing a quite large range of parameterizations, we aim at identifying possible links with general properties of the EOS, such as the symmetry energy or the effective mass,   and at isolating, if any,
features which might be connected to the specific properties of our LYE-inspired parameterizations.

In the case of the heaviest isotopes, Zr and Pb (panels (c) and (d)), a well pronounced neutron skin shows up in very asymmetric nuclei. Such a skin does not seem to be affected by the 
typical very low-density behavior of the YGLO functional since one may easily see that the two YGLO predictions are 
 systematically located between other Skyrme EDFs curves. 

One may easily recognize, on the other hand, that the trend reflects the known correlation existing between the neutron skin thickness and the slope $L$ of the symmetry energy calculated at the saturation density, which was discussed for instance in Refs. \cite{centelles,warda}. The ordering encountered in the neutron skin thickness predictions of the most neutron-rich nuclei exactly reflects the one observed for $L$ at the saturation density (Fig. \ref{densin}, panel (c)). 
The explanation for this trend is known: in the central region of these neutron-rich nuclei, where the total density is close to the saturation density, there is a neutron excess. A larger value of $L$ indicates a stiffer NM EOS, which means that extra neutrons tend to become more unbound more rapidly around the saturation density. There is then a stronger tendency to push neutrons outside and to form a neutron skin.

It should be worthwhile to notice that some recent works based on phenomenological relativistic EDF have highlighted that the correlation between the neutron-skin thickness \cite{typel} and the slope of the nuclear symmetry energy is however modified when $\alpha$-cluster formation, which is not implemented within the adopted theoretical framework, is taken into account. 

In the case of the lightest systems (O and Ca in panels (a) and (b)), the ordering of the results for neutron-rich nuclei reflects once again the one already found for $L$ around saturation. 
However, in this case, the differences between the results obtained with the considered parametrizations are less pronounced. This 
occurs because the corresponding values of $L$ at saturation are quite close.

It is interesting to analyse the behavior of neutron-deficient nuclei in O and Ca isotopes.  In the case of $^{12}$O and $^{34}$Ca, a quite large difference between neutron and proton radii is predicted, around 0.5 and 0.3 fm, respectively, with a remarkable spread of values depending on the employed functional. To the best of our knowledge, any overt correlation between general properties of the EOS and such a quantity displayed in Fig. \ref{skin} (which might be related to the development of a so-called proton skin) has ever been discussed in the literature. In the next section, we propose, for the first time, a possible correlation, at least from a qualitative point of view, with the isoscalar effective mass.

\section{Tails of density profiles: links with the effective mass and the slope of the symmetry energy at low density}
\label{tail}

With the aim to analyze in more detail possible connections between surface properties and the low-density behavior associated with the YGLO functionals, we 
study in this section neutron and proton density profiles, plotted in logarithmic scale, for selected neutron-rich and neutron-deficient isotopes.

\begin{figure*}[t]
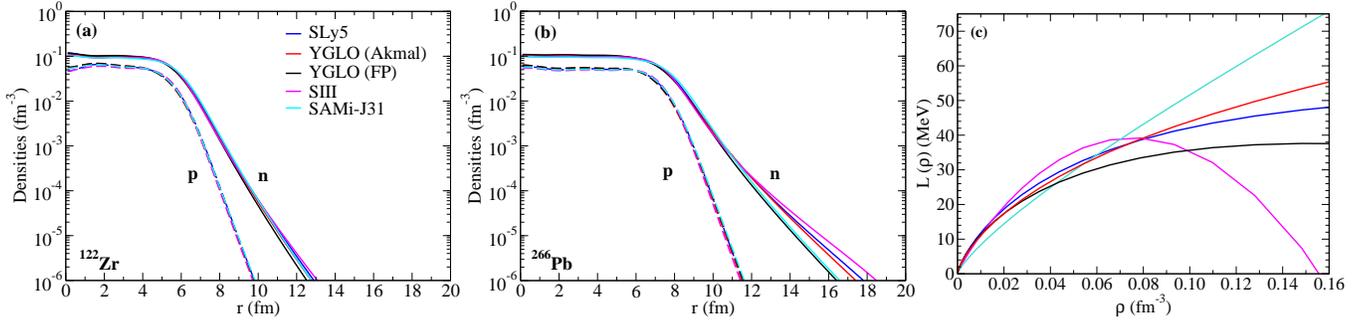

\begin{center}
\includegraphics[width=0.7\columnwidth]{Zr_nr.eps}\includegraphics[width=0.7\columnwidth]{Pb_nr.eps}\includegraphics[width=0.66\columnwidth]{slope.eps}
\caption{Density profiles of neutrons and protons in logarithmic scale for $^{122}$Zr (a) and $^{266}$Pb (b) and density behavior of the slope of the symmetry energy (c), for different EDFs considered in our study.}
\label{densin}
\end{center}
\end{figure*}

We show in Fig. \ref{densin} these profiles for 
two neutron-rich nuclei, 
$^{122}$Zr (a) and $^{266}$Pb (b), where the most diffuse densities are the neutron ones. In addition to the YGLO and SLy5 curves and in analogy with what we have done in Fig. \ref{skin}, we also plot the densities obtained with SIII and SAMi-J31, in order to employ EDFs which differ quite strongly in the stiffness of their NM EOSs. 

\begin{figure*}[t]
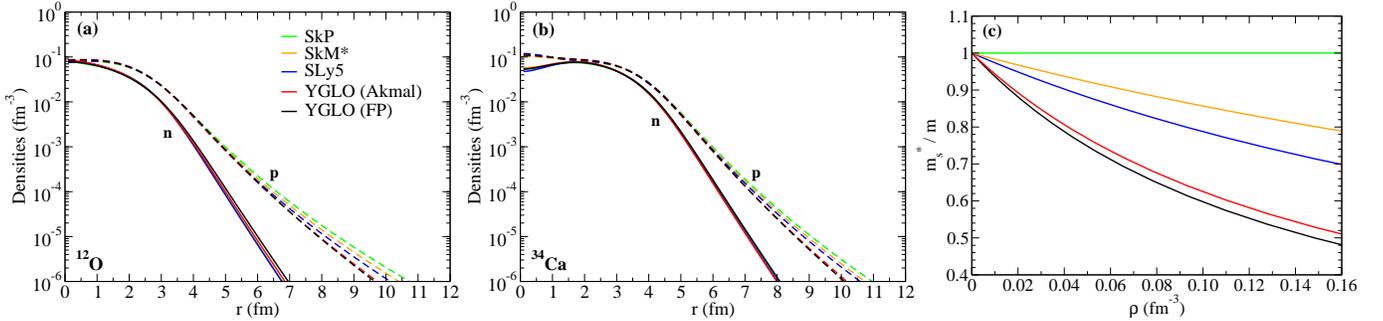

\begin{center}
\includegraphics[width=0.7\columnwidth]{O_nd.eps}\includegraphics[width=0.7\columnwidth]{Ca_nd.eps}\includegraphics[width=0.68\columnwidth]{eff_mass.eps}
\caption{Density profiles of neutrons and protons in logarithmic scale for $^{12}$O (a) and $^{34}$Ca (b) and density behavior of the isoscalar effective mass (c), for different EDFs considered in our study.}
\label{densip}
\end{center}
\end{figure*}

We observe that, in the tails of the neutron densities, some differences start to become visible around 10$^{-3}$-10$^{-4}$ fm$^{-3}$. These differences are not qualitatively coherent with  those previously observed for the neutron skin thicknesses in Fig. \ref{skin} since the ordering is not the same. 

It is easy to realize that the differences found in the neutron low-density tails still depend on the slope of the symmetry energy. However, this time, such a slope is not the one computed at saturation density but the one computed at lower densities.
One may see for example in panel (c) of Fig. \ref{densin} that, at densities lower than $\approx$ 0.05 fm$^{-3}$, the SIII NM EOS starts to be the stiffest EOS, whereas both YGLO NM EOSs are less stiff than the SLy5 NM EOS. 
The interpretation is the same as the one already provided for neutron skin thicknesses. A larger value of $L$ at densities 
$\approx$ 0.05 fm$^{-1}$ implies there a stronger tendency of pushing neutrons outside at lower densities, that is in the tails of the density profiles. 

The behavior in this density region cannot be however regarded as representative of the YGLO functional, 
that is, related to the very low-density behavior driven by the LYE. 
Indeed, the two YGLO parameterizations do not predict the most compact tails among the considered EDFs. Nevertheless, this behavior reflects a peculiar trend in  a density regime where the NM YGLO EOSs were benchmarked on ab-initio results.
For this reason, such YGLO predictions should be regarded as more reliable compared to those obtained with more traditional EDFs.

Looking at the lightest isotopic chains (Oxygen and Calcium) in Fig. \ref{skin}, one observes that both YGLO functionals provide more compact neutron-deficient systems than those predicted with the other EDFs. The difference of the neutron and proton radii is indeed found to be smaller with the YGLO functionals, indicating that proton surfaces are expected to be less diffuse with YGLO.   
We plot in Fig. \ref{densip} the proton and neutron density profiles, in logarithmic scale, for two neutron-deficient systems, $^{12}$O and $^{34}$Ca. 
This time, the external profiles are the proton ones.
In addition to the YGLO and SLy5 cases, the density profiles are plotted also for two different functionals, SkM*  and SkP, having quite different isoscalar effective masses.  Effective masses are equal to 0.468, 0.497, 0.697, 0.789, and 1.000 at saturation 
for YGLO(FP), YGLO(Akmal),  SLy5, SkM*, and SkP, respectively (see Fig. \ref{densip}, panel (c)).
The differences which exist in the low-density tails of the proton profiles (panels (a) and (b) of Fig. \ref{densip}) reflect this time the same ordering as the one observed in the differences between neutron and proton radii. 

We have identified the existence of a correlation of these spatial properties with the isoscalar effective mass of infinite matter.
In the same way as, for neutron-rich systems, radii differences and tails were connected to $L$ at different densities (at saturation for the radii and at lower densities for the tails), we may now connect radii differences and tails to $m^*_s/m$. Lower effective masses around saturation would imply more compact proton radii, whereas lower effective masses at lower densities would imply less extended tails. Since the different effective mass curves do not cross each other when the density increases (panel (c) of Fig. 9), the ordering is now found to be the same for both quantities.

Therefore,  whereas spatial surface features for neutron-rich systems (with a quite large isospin asymmetry) have been related to an isovector property (the slope of the symmetry energy), they have been found to be driven by the isoscalar effective mass 
in neutron-deficient nuclei (where the isospin asymmetry is much smaller). 
These correlations might deserve a deeper investigation. Some recent works have established the existence of a close relationship between
low-lying isoscalar dipole modes in nuclei and surface properties in the corresponding density profiles, independently of 
the isospin asymmetry of the system \cite{zheng, bur19, burFro}. Both neutron-rich and neutron-deficient systems that we have analyzed  in Fig. \ref{densip} have a quite pronounced and diffuse surface. In view of all these results, we might thus expect that a link could finally exist between low-lying isoscalar dipole excitations in nuclei with a diffuse surface and isoscalar effective masses (slopes of the symmetry energy at low densities) in neutron-deficient (neutron-rich) systems. We reserve this investigation for a future study.

%
\section{Effective mass} \label{emass}
%
As already underlined several times, the YGLO functionals which have been introduced so far have very low 
isoscalar and neutron effective masses. We have seen for example that this has a visible impact on shell closures, which have been found to be much more pronounced compared to the experimental data and to predictions obtained  with a traditional Skyrme functional. 

\begin{figure}[t]
\begin{center}
\includegraphics[width=\columnwidth]{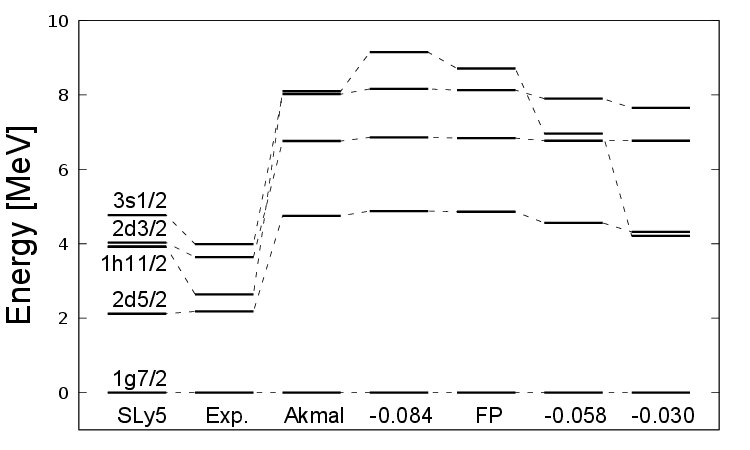}
\vspace*{-0.7cm}
\caption{Single-particle energies of the last occupied proton states in the nucleus $^{208}$Pb, where the 1g7/2 state is used as reference zero energy in each case. The experimental values are extracted from Ref. \cite{sch}. SLy5 predictions are shown for comparison. Akmal and FP correspond to the two YGLO functionals where the optimal $W$ value is used in each case. The other shown results correspond to the YGLO(FP) functional, where $W$ is varied taking the values -0.084, -0.058, and -0.030. 
}
\label{levels}
\end{center}
\end{figure}

To visualize the effect of such low effective masses on the level densities, we show as an illustration in Fig. \ref{levels} the energies of the last occupied proton states in the nucleus $^{208}$Pb. The experimental values are extracted from Ref. \cite{sch}. The single-particle energies of the 1$g_{7/2}$ state are used as reference zero energy in all cases. For comparison with a Skyrme case, also the SLy5 results are plotted. For the YGLO functional, we show the Akmal and FP energies, using in each case the associated optimal $W$ value, as well as the FP energies obtained with three different values of $W$, -0.084, -0.058, and -0.030. In such a way, the four plotted YGLO(FP) results correspond to increasing values of the isoscalar effective mass going from the left to the right in the figure, $W=-0.058$ and $W=-0.030$ corresponding in particular to $m_s^*/m=$ 0.55 and $m_s^*/m=$ 0.70, respectively. We observe a trend which is coherent with what could have been expected. In general, all the YGLO functionals lead to a less compact single-particle spectrum owing to the low isoscalar effective mass (one can see in particular the large gap between the 1$g_{7/2}$ and 2$d_{5/2}$ energies). If one now considers the four FP cases, one may notice that, moving from the left to the right in the figure, the level density is increased (even if still not enough compared to the experimental spacing of single-particle states), which is indeed the expected trend. 

\begin{figure}[t]
\begin{center}
\includegraphics[width=\columnwidth]{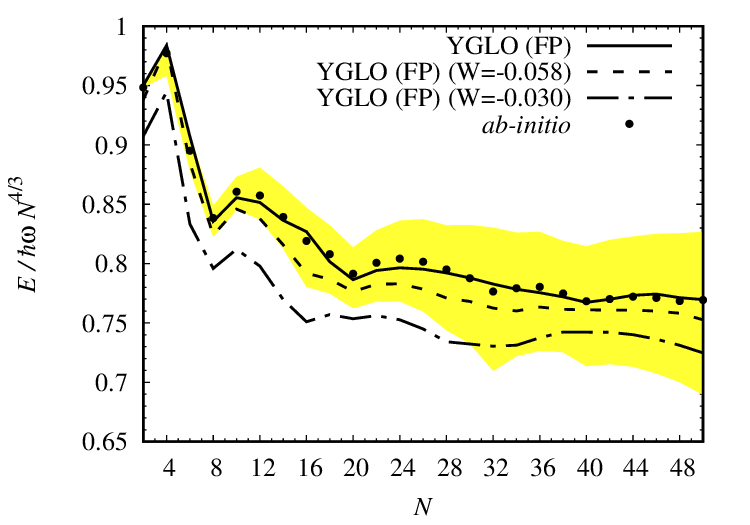}
\vspace*{-0.7cm}
\caption{Drop energies for the trap frequency $\hbar \omega$ = 10 MeV. The yellow area and the circles are the same as in Fig. \ref{dr}. The solid line represents the energies obtained using the YGLO(FP) parametrization introduced in this work. The other two lines correspond to the same parametrization, where only the value of $W$ is changed to -0.058 (dashed line) and -0.030 (dot-dashed line).}
\label{drops_effmass}
\end{center}
\end{figure}

Figure \ref{drops_effmass} shows the effect of varying $W$ on the energies of neutron drops with a trap frequency of 10 MeV. Together with the FP parametrization introduced in this work, we plot also the FP curves obtained with the two $W$ values  -0.058 and -0.030, while maintaining constant all the other parameters. One may see that the agreement with the benchmark curve is deteriorated with these two choices of $W$  (that is, with the increase of the effective mass). It is worthwhile to notice that such a deterioration of the agreement with the benchmark curve would be observed also if a global adjustment including pairing and spin-orbit coupling constants was carried out, with the constraint of having an isoscalar effective mass higher than 0.5.

\begin{figure}[t]
\begin{center}
\includegraphics[width=\columnwidth]{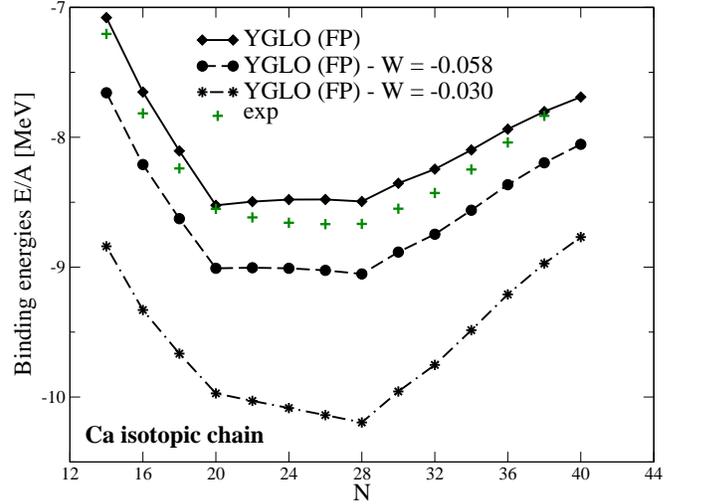}
\vspace*{-0.7cm}
\caption{Binding energies per nucleon for the Ca isotopic chains. The same three FP parametrizations as in Fig. \ref{drops_effmass} are used and the experimental data as in Fig. \ref{be} are also reported.}
\label{nu_effmass}
\end{center}
\end{figure}

The observed differences reflect also in the binding energies per nucleon, which are shown in Fig.\ref{nu_effmass} for the Ca isotopic chain,  using the same three FP parametrizations as in Fig. \ref{drops_effmass}. Also in this case the change of $W$, which implies an increase of the effective mass, leads to overbound systems. 

It seems that a higher effective mass is not compatible with correct energies in neutron drops and, also, leads to overbound finite nuclei in the illustrative case of the Ca isotopic chain.  At the same time, effective masses as low as we  have in the YGLO case are  not acceptable because they lead to level densities which are too low around the Fermi energy and to overpronounced shell closures.

It was shown in Ref. \cite{drop2} that the inclusion of the $p$-wave contribution in the ELYO functional allowed for obtaining both satisfactory energies in neutron drops and a higher effective mass compared to the YGLO case. This would indicate that, also in the YGLO case, a promising direction to follow would be to include the $t_2$-like contribution in the functional (so far, such a contribution was set equal to zero for simplicity) and to adjust again 
on 
ab-initio
drop energies 
to design a functional that can be used for finite-size systems. 

We stress that, obviously, the EOSs of matter would not be affected at all by this procedure. On the other side, the hope would be to obtain, as in the ELYO case, a satisfactory reproduction of ab-initio benchmark energies in neutron drops and a reasonable value for the effective mass. Such a functional could then be checked on nuclei. We expect that the quality of the predictions should be improved since the effective mass would be higher. 

We will then be able to draw our final conclusions and state whether a fine tuning on observables of a few nuclei is necessary or not for this functional. If it turns out that a fine tuning is required, this would imply that the ab-initio content of the YGLO functional would be attenuated compared to its present form. We remind however that also several chiral potentials have been fine tuned on nuclei to improve for example their description of nuclear binding energies and/or radii (see for instance Ref. \cite{ek}).  

%
\section{Conclusions and perspectives} \label{concl}
%
%
We have applied for the first time the YGLO functionals to ground-state properties of finite nuclei. 
First, the Akmal version of such a functional has been extended to neutron drops and a parametrization has been introduced for this case. A new alternative parametrization has been introduced in neutron drops also for the FP case. This parametrization has been used in practice for the applications carried out here for finite nuclei, because it leads in general to nuclear binding energies closer to the experimental ones (compared to the previous version). 

Two-neutron separation and binding energies have been analyzed first and compared with experimental values and Skyrme SLy5 predictions. Whereas nuclear separation energies are reasonably well reproduced, we have shown that YGLO functionals describe less well the absolute binding energies, the Akmal version working better than the FP one.  

On the other side, charge radii are correctly predicted by the YGLO functionals as shown by the comparison with the experimental values of a set of charge radii in a selection of closed-shell nuclei. 

Differences of neutron and proton radii and tails of density profiles have been also discussed. In the neutron-rich systems under study, it has been shown that both the neutron skin thicknesses and the very low-density tails of neutron density profiles depend on the slope $L$ of the symmetry energy around saturation and at lower densities, respectively. Since the $L$ curves for the various functionals used here cross each other when the density is varied, neutron skin thicknesses and neutron density tails do not present the same ordering of the results, according to the used functional. In neutron-deficient nuclei, on the other side, these two spatial properties have been shown to be correlated to the isoscalar effective mass. Since the effective mass curves for the various functionals employed here do not cross each other when the density increases, the same ordering of the results (according to the used functional) has been found in both radii differences and tails of proton densities. We plan to further investigate this aspect by computing in a future work isoscalar dipole low-energy excitations in neutron-rich (neutron-deficient) nuclei, to check whether the found correlations with the slope of the symmetry energy (the isoscalar effective mass) may be visible also in such low-lying excitations, in view of their established strong sensitivity to spatial surface properties. 

We have found in general that the peculiar behavior of YGLO functionals at extremely low densities, which is driven by the LYE, does not affect in any distinctive or systematic way the ground-state properties that we have analyzed. Nevertheless, the behavior shown by the tails of neutron and proton density profiles reflects the characteristic trend exhibited in a density regime where the YGLO predictions, benchmarked on ab-initio calculations, might be considered more reliable than the traditional EDFs one. 

Finally, effective masses have been discussed. They are low in YGLO functionals and this affects shell closures and level densities around the Fermi energy. However, we have illustrated that changing the splitting parameter $W$ to have a higher and more acceptable effective mass induces a deterioration in the predictions of energies in both neutron drops and finite nuclei. It seems that, for the present form of these functionals, a simultaneous adjustment of energies and effective masses is not feasible. A possible direction to improve this aspect will be to enlarge the form of the functional, by including also a $p$-wave $t_2$ contribution. This is reserved for a future work. 

%
\section*{Acknowledgment} 
M.G. aknowledges funding from the European Union's Horizon 2020 research and innovation program under Grant Agreement No. 654002 and from the IN2P3-CNRS BRIDGE-EDF 
project.  


\begin{thebibliography}{99}
%

\bibitem{f1} R. J. Furnstahl, in {\it{Renormalization Group and Effective Field Theory Approaches to Many-Body Systems}}, edited by A. Schwenk and J. Polonyi, Lecture Notes in Physics, Vol. 852 (Springer, Berlin, 2012), Chap. 3. 
\bibitem{fur} R. J. Furnstahl, Eur. Phys. J. A 56, 85 (2020).
\bibitem{ppnp} M. Grasso, Prog. Part. Nucl. Phys. 106, 256 (2019). 
\bibitem{yang2017} C. J. Yang, M. Grasso, and D. Lacroix, Phys. Rev. C 96, 034318 (2017).
\bibitem{burrello2020} S. Burrello, M. Grasso, and C. J. Yang, Phys. Lett. B, 135938 (2020).
\bibitem{yglo} C. J. Yang, M. Grasso, and D. Lacroix, Phys. Rev. C 94, 031301(R) (2016).
\bibitem{elyo} M. Grasso, D. Lacroix, and C. J. Yang, Phys. Rev. C 95, 054327 (2017).
\bibitem{drop1} J. Bonnard, M. Grasso, and D. Lacroix, Phys. Rev. C 98, 034319 (2018); Phys. Rev. C 103, 039901(E) (2021).
\bibitem{drop2} J. Bonnard, M. Grasso, and D. Lacroix, Phys. Rev. C 101, 064319 (2020).
\bibitem{LY} T. D. Lee and C. N. Yang, Phys. Rev. 105, 1119 (1957).
\bibitem{HY} K. Huang and C. N. Yang, Phys. Rev. 105, 767 (1957). 
\bibitem{efimov} V. N. Efimov, M. Ya. Amusia, Sov. Phys. JETP 20, 388 (1965).  
\bibitem{baker} G. A. Baker, Rev. Mod. Phys. 43, 479 (1971). 
\bibitem{bishop} R. F. Bishop, Ann. Phys. 77, 106 (1973).
\bibitem{ya} M. Ya. Amusia and V. N. Efimov, Ann. Phys. (NY) 47, 377 (1968). 
\bibitem{HF} H. W. Hammer and R. J. Furnstahl, Nucl. Phys. A 678, 277 (2000).
\bibitem{sc} C. Wellenhofer, C. Drischler, and A. Schwenk, Phys. Lett. B 802, 135247 (2020). 
\bibitem{schafer} T. Sch\"afer, C.-W. Kao, S.R. Cotanch, Nucl. Phys. A 762, 82 (2005).
\bibitem{kaiser} N. Kaiser, Nucl. Phys. A 860, 41 (2011).
\bibitem{steele} J.V. Steele, arXiv:nucl-th/0010066v2
\bibitem{FW} A. L. Fetter and J. D. Walecka, {\it{Quantum Theory of Many-Particle Systems}} (McGraw-Hill, New York, 1971).
\bibitem{bur15} S. Burrello, F. Gulminelli, F. Aymard, M. Colonna, and Ad. R. Raduta, Phys. Rev. C 92, 055804 (2015).
\bibitem{sk1} T. H. R. Skyrme, Phil. Mag. 1, 1043 (1956).
\bibitem{sk2} T. H. R. Skyrme, Nucl. Phys. 9, 615 (1959).
\bibitem{vau} D. Vautherin and D. M. Brink, Phys. Rev. C 5, 626 (1972).
\bibitem{negele} J.W. Negele and D. Vautherin, Nucl. Phys. A 298 (1973). 
\bibitem{girod} M. Girod and P. Schuck, Phys. Rev. Lett. 111, 132503 (2013). 
\bibitem{gogny1} D. Gogny, Nucl. Phys. A 237, 399 (1975).
\bibitem{gogny2} D. Gogny, in: G. Ripka G, M. Porneuf (Eds.), Proceedings of the International Conference on Nuclear Selfconsistent Fields, Trieste 1975, North Holland, Amsterdam, 1975. 
\bibitem{guarnera} A. Guarnera, M. Colonna, P. Chomaz, Phys. Lett. B 373, 267 (1996).
\bibitem{geze} A. Gezerlis and J. Carlson, Phys. Rev. C 81, 025803 (2010). 
\bibitem{akmal} A. Akmal, V. R. Pandharipande and D. G. Ravenhall, Phys. Rev. C 58, 1804 (1998).
\bibitem{FP} B. Friedman and V. Pandharipande, Nucl. Phys. A 361, 502 (1981).
\bibitem{maris} P. Maris, J.P. Vary, S. Gandolfi, J. Carlson, and S.C. Pieper, Phys. Rev. C 87, 054318 (2013). 
\bibitem{gandolfi} S. Gandolfi, J. Carlson, and S.C. Pieper, Phys. Rev. Lett. 106, 012501 (2011). 
\bibitem{potter} H.D. Potter, S. Fischer, P. Maris, J.P. Vary, S. Binder, A. Calci, J. Langhammer, and R. Roth, Phys. Lett. B 739, 445 (2014). 
\bibitem{nndc}{From ENSDF database as of October 7$^{th}$, 2020. Version available at http://www.nndc.bnl.gov/ensarchivals/}
\bibitem{sly5} E. Chabanat, P. Bonche, P. Haensel, J. Meyer, R. Schaeffer, Nucl. Phys. A 627, 710 (1997); 635, 231 (1998); 643, 441 (1998).
\bibitem{gil} H. Gil, P. Papakonstantinou, C.H. Hyun, and Y. Oh, Phys. Rev. C 99, 064319 (2019). 
\bibitem{raexp} http://nrv.jinr.ru/nrv/ 
\bibitem{skm} J. Bartel, P. Quentin, M. Brack, C. Guet, H.-B. Hakansson, Nucl. Phys. A 386, 79 (1982).
\bibitem{skp} J. Dobaczewski, H. Flocard, J. Treiner, Nucl. Phys. A 422, 103 (1984).
\bibitem{siii} M. Beiner, H. Flocard, Nguyen Van Giai, and P. Quentin, Nucl. Phys. A 238, 29 (1975).
\bibitem{sami} X. Roca-Maza, M. Brenna, B. K. Agrawal, P. F. Bortignon, G. Col\`o, L. G. Cao, N. Paar, and D. Vretenar, Phys. Rev. C 87, 034301 (2013).
\bibitem{centelles} M. Centelles, X. Roca-Maza, X. Vi\~nas, and M. Warda, Phys. Rev. Lett. 102, 122502 (2009).
\bibitem{warda} M. Warda, X. Vi\~nas, X. Roca-Maza, and M. Centelles, Phys. Rev. C 80, 024316 (2009). 
\bibitem{typel} S. Typel, Phys. Rev. C 89, 064321 (2014). 
\bibitem{zheng} H. Zheng, S. Burrello, M. Colonna, and V. Baran, Phys. Rev. C 94, 014313 (2016).
\bibitem{bur19} S. Burrello, M. Colonna, G. Col\`o, D. Lacroix, X. Roca-Maza, G. Scamps, and H. Zheng, Phys. Rev. C 99, 054314 (2019).
\bibitem{burFro} S. Burrello, M. Colonna, and H. Zheng, Front. Phys. 7,  53  (2019).
\bibitem{sch} N. Schwierz, I. Wiedenhover, and A. Volta, arXiv:0709.3525.
\bibitem{ek} A. Ekstr\"om et al., Phys. Rev. C 91, 051301(R) (2015). 

%
\end{thebibliography}
\end{document}